\documentclass[aps,prb,twocolumn,superscriptaddress]{revtex4-2}
\usepackage{bm}
\usepackage{graphicx}
\usepackage{mathrsfs}
\usepackage{amsmath,amssymb}

\begin{document}
\title{Coherent exciton spin dynamics and three-dimensional quantum state tomography \\ 
in a single InAlAs quantum dot}

\author{J.\ Njala}
\affiliation{Graduate School of Engineering, Hokkaido University, N13 W8, Kitaku, Sapporo 060-8628, Japan}

\author{Y.\ Yamamoto}
\affiliation{Graduate School of Engineering, Hokkaido University, N13 W8, Kitaku, Sapporo 060-8628, Japan}

\author {R.\ Kaji}
\affiliation{Graduate School of Engineering, Hokkaido University, N13 W8, Kitaku, Sapporo 060-8628, Japan}

\author{S.\ Adachi}
\affiliation{Graduate School of Engineering, Hokkaido University, N13 W8, Kitaku, Sapporo 060-8628, Japan}

\author{H.\ Sasakura}
\email{Contact author: hirotaka@eng.hokudai.ac.jp}
\affiliation{Graduate School of Engineering, Hokkaido University, N13 W8, Kitaku, Sapporo 060-8628, Japan}

\date{\today}

\begin{abstract}
We investigate the coherent exciton spin dynamics in a single InAlAs/AlGaAs quantum dot using time-resolved quantum state tomography. Under two-LO-phonon quasi-resonant excitation of neutral exciton, we observe pronounced quantum beats in the circular and diagonal polarization components, reflecting the fine-structure splitting ($\Delta \approx 19.6 \ \mu\text{eV}$). By employing a global fitting procedure across three orthogonal polarization bases, we demonstrate that the spin evolution is consistently described by a unified Hamiltonian dominated by the anisotropic exchange interaction. While the initial degree of circular polarization is limited to $\approx 0.28$ due to fast relaxation processes during carrier cooling, the subsequent dynamics reveal a long-lived spin coherence ($1.1 \pm 0.2$ ns) that exceeds the exciton lifetime ($\sim 767$ ps). Our analysis reveals that the spin-formation time is significantly shorter than the instrument response function, and the absence of a discernible Overhauser shift confirms a negligible influence from the local nuclear environment under the present conditions. These results provide a quantitative benchmark for the three-dimensional reconstruction of spin trajectories using differential polarization signals, demonstrating the feasibility of using quasi-resonant excitation for stable spin initialization in semiconductor nanostructures.

\end{abstract}

\maketitle

\section {Introduction}

Semiconductor quantum dots (QDs) have emerged as a promising platform for quantum information processing due to their ability to host stable, optically addressable exciton spins~\cite{book1,shie07}. The coherent manipulation of these spins is fundamental for various applications, ranging from high-speed quantum gates~\cite{Press08,Ber08} to photonic interfaces~\cite{DeGreve12,Lodahl15}. The spin evolution of a neutral exciton is intrinsically governed by the anisotropic exchange interaction (AEI) arising from the lowered $C_{2v}$ symmetry of the quantum dot structure~\cite{Gammon96, Bayer02, Tar04}. This interaction results in the fine-structure splitting (FSS) of the bright exciton states into two linearly polarized components~\cite{Ivchenko97, Flis01}. 

While the fundamental physics of these quantum beats was established in earlier studies~\cite{Senes05}, most experiments have relied on a single observation basis, primarily the degree of circular polarization. However, a complete understanding of the three-dimensional spin trajectory requires simultaneous verification across orthogonal bases. Such a comprehensive approach, known as quantum state tomography, has recently been developed to characterize light-matter interactions in single QD-cavity systems~\cite{Carl17} and to reconstruct the full density matrix of QD spin qubits, such as  the dark exciton state~\cite{Cog20}.

 This methodology is essential to disentangle the bright exciton dynamics originating from exchange interactions from other physical processes, such as the hyperfine interaction with the underlying nuclear spin ensemble~\cite{Brau05, Urb13} or non-equilibrium carrier relaxation processes. While the methodology for full spin-state reconstruction is well-established in ensemble studies~\cite{Kik98,Kosa08}, its rigorous application to the fast dynamics of a single QD highlights the precision of this classic approach in the quasi-resonant regime.

In this work, we present a detailed study of the coherent exciton spin evolution in a single InAlAs/AlGaAs quantum dot through high-resolution time-resolved quantum state tomography. By analyzing the time evolution of the differential signals $D_i$ in three orthogonal bases [$i=\text{circ,\ (circular),\  diag\ (diagonal),\ rect\ (rectilinear)}$], we achieve a quantitative reconstruction of the spin trajectory on the Bloch sphere. By employing two-LO-phonon quasi-resonant excitation, we achieve well-defined spin initialization and observe pronounced quantum beats with a fine-structure splitting of $\Delta \approx 19.6 \ \mu\text{eV}$. Rather than requiring models that incorporate complex spin-formation delays or significant nuclear-induced tilting of the precession axis, our results demonstrate that the spin dynamics are quantitatively reconstructed by a unified AEI Hamiltonian using a global fitting procedure across the circular, diagonal, and rectilinear bases. Furthermore, this tomographic approach allows us to disentangle the intrinsic spin-dephasing time from the exciton lifetime, revealing a long-lived coherence of the reconstructed three-dimensional spin vector $|\boldsymbol{S}(t)|$ that extends beyond the carrier lifetime. The consistency of our results with the theoretical model, which aligns with recent advancements in the field~\cite{Gio25}, provides a reliable benchmark for the verification and control of spin qubits in epitaxial nanostructures.

\section {EXPERIMENTS}
The sample studied in this work consists of self-assembled $\text{In}_{0.75}\text{Al}_{0.25}\text{As/Al}_{0.3}\text{Ga}_{0.7}\text{As}$ QDs grown on an undoped (100)-GaAs substrate by molecular beam epitaxy~\cite{Sa04}. Based on atomic force microscopy of a reference uncapped QD layer, the QDs were characterized by an average diameter of $\sim$20 nm, a height of $\sim$4 nm, and a density of $\sim$$5\times 10^{10} \text{ cm}^{-2}$. To isolate and address individual QDs, small mesa structures were fabricated using electron-beam lithography and wet etching.

Micro-photoluminescence ($\mu$-PL) measurements were performed at 20 K using a closed-cycle helium cryostat. For spin initialization, the QD was excited by left-circularly polarized ($|L\rangle$) pulses from a mode-locked Ti:sapphire laser (repetition rate: 76 MHz, pulse width: $\sim$3 ps). The excitation wavelength was tuned to $\sim$752 nm, corresponding to the two-LO-phonon resonance of the neutral exciton ($X^0$). This quasi-resonant excitation facilitates stable spin initialization with an initial polarization degree sufficient to resolve the coherent dynamics, while suppressing carrier-induced decoherence and random capture processes typically associated with non-resonant excitation into the wetting layer or barrier.

The QD emissions were collected and analyzed using a triple-grating spectrometer equipped with a Si-charge coupled device camera, providing a spectral resolution of $\sim$7 $\mu$eV. This setup facilitated a high-precision analysis of the emission; by performing Lorentzian line-shape fitting, the central energy of the peaks could be determined with an accuracy of $\sim$1 $\mu$eV. Figure 1(a) shows a measured $\mu$-PL spectrum under quasi-resonant excitation. To identify the various exciton complexes, we investigated the linear polarization dependence by using a rotating half-wave plate (HWP) and a fixed linear polarizer inserted into the detection path~\cite{Ad07}. As shown in Fig. 1(b), the peak energy of the line labeled $X^{0}$ exhibits a clear sinusoidal oscillation reflecting the AEI. The experimental data were well-fitted by a sine function, yielding a FSS of $\Delta = 19.8 \pm 0.4 \ \mu$eV. The HWP angles corresponding to the maximum and minimum energies identify the orientation of the $|H\rangle$ and $|V\rangle$ excitonic eigenstates, which define the internal reference frame for the subsequent time-resolved quantum state tomography.


\begin{figure}
\centering
\includegraphics[width=0.9\columnwidth]{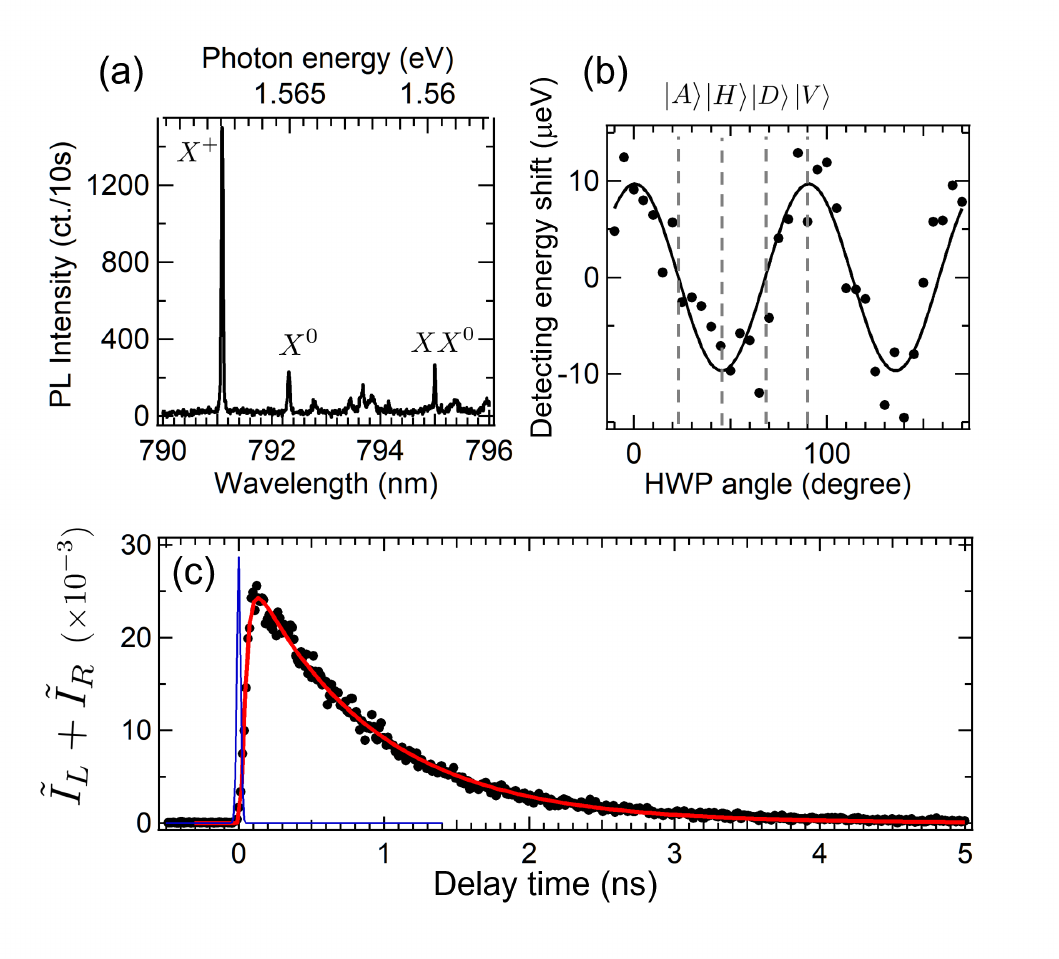}%
\caption{\label{fig1} Static photoluminescence characterization and temporal evolution of the neutral exciton. (a) $\mu$-PL spectrum of a single InAlAs/AlGaAs QD under two-LO-phonon quasi-resonant excitation at 20 K. The labels $X^+$, $X^0$, and $XX^0$ denote the ground state emissions of the positively charged exciton, neutral exciton, and neutral biexciton, respectively. While the $X^+$ emission exhibits significant intensity, the $X^0$ peak is well-resolved spectrally, allowing for the selective initialization and observation of the neutral excitonic ground state via the phonon-assisted process. (b) Exciton peak energy shift as a function of the HWP angle in the detection path. The solid line represents a sinusoidal fit, yielding a FSS of $\Delta = 19.8 \pm 0.4~\mu\text{eV}$. Vertical dashed lines indicate the HWP angles corresponding to the excitonic energy eigenstates ($|H\rangle$ and $|V\rangle$) and their diagonal superposition states ($|D\rangle$ and $|A\rangle$). These orientations define the laboratory reference frames required for the full three-dimensional quantum state tomography. (c) TRPL signal $\tilde{I}_L + \tilde{I}_R$ for the $X^0$ emission. The blue profile represents the IRF, and the red solid line shows a single-exponential fit convolved with the IRF. The intrinsic exciton lifetime, extracted by deconvolving the IRF from the TRPL signal, is $\tau =767\pm 10~\text{ps}$. The near-instantaneous rise of the emission within the IRF limit highlights the rapid carrier injection achieved by the two-LO-phonon quasi-resonant excitation.
}
\end{figure}

For the time-resolved PL (TRPL) measurements, a narrow-bandpass filter with a 0.5-nm bandwidth was employed to selectively isolate the $X^0$ emission. Polarization-sensitive detection was implemented using a motorized analysis stage equipped with a quarter-wave plate, a HWP, and a linear polarizer. The time-dependent PL signals were recorded using a time-correlated single-photon counting (TCSPC) system. The overall instrument response function (IRF) of the system was $\sim$30 ps, determined by the jitter of the single-photon detector and the laser pulse width. 

To achieve full quantum state tomography, the PL intensities were sequentially measured in the circular ($|L\rangle/|R\rangle$), diagonal ($|D\rangle/|A\rangle$), and rectilinear ($|H\rangle/|V\rangle$) bases. Specifically, the three Stokes parameters were extracted from these six polarization components to reconstruct the time-dependent spin vector on the Bloch sphere (the quantum-mechanical correspondence of the Poincar\'e sphere). To compensate for technical imbalances in the detection efficiencies across different channels, the integrated intensities of each polarization component were used to normalize the respective time-dependent traces. We then deduce the differential signals $D_{i}$ that faithfully represent the relative components of the spin vector, providing a reliable basis for the subsequent three-dimensional reconstruction of the spin trajectory $|\boldsymbol{S}(t)|$ and the global fitting analysis.

\section{Results and Discussion}
The TRPL signals, recorded as coincidence counts between the $X^0$ emission and the laser trigger, provide a direct visualization of the exciton spin dynamics. We first determine the exciton lifetime by analyzing the sum signal $\tilde{I}_{L} + \tilde{I}_{R}$, which represents the total population of the $X^0$ state as shown in Fig. 1(c). Here, $\tilde{I}_{j}$ denotes the normalized transient signal to compensate for channel-dependent detection efficiencies and minor optical misalignments, defined as $\tilde{I}_{j} = I_j(t) / \int I_j(t)dt$. While we initially focus on the circular basis ($j \in \{L, R\}$) for population analysis, this normalization is applied consistently to all measurement channels ($j \in \{L, R, D, A, H, V\}$) for the subsequent tomographic reconstruction.

The data were fitted with a model function convolved with the IRF, $f(t) \ast g(t)$, where $f(t) = f_{0}\exp(-t/\tau)(1-\exp(-t/\tau_{p}))$ accounts for the population decay and the carrier-filling process, and $g(t)$ is a Gaussian representing the IRF (FWHM $\sim 30$ ps). This procedure yielded an intrinsic exciton lifetime of $\tau = 767 \pm 10$ ps and a population rise time of $\tau_{p} = 34 \pm 2$ ps, with the initial amplitude $f_{0} = (28.6 \pm 0.2) \times 10^{-3}$, which serves as the normalization base for evaluating the quantum beat contrast. The rapid rise of the signal, nearly limited by the IRF, confirms the efficient and near-instantaneous carrier injection achieved by the two-LO-phonon quasi-resonant excitation.

To quantify the three-dimensional spin trajectory, we define the time-resolved differential signals $D_i$ ($i = \text{circ, diag, rect}$) based on the normalized intensities introduced above. Specifically, the polarization dynamics are extracted as the differences between these signals: $D_{\rm circ} = \tilde{I}_L - \tilde{I}_R$, $D_{\rm diag} = \tilde{I}_A - \tilde{I}_D$, and $D_{\rm rect} = \tilde{I}_H - \tilde{I}_V$. This approach not only corrects for instrumental biases but also suppresses the amplification of experimental noise in the low-count regime. By utilizing time-integrated intensities for normalization rather than instantaneous total counts, we avoid the numerical instability and large fluctuations that typically occur as the PL signal decays, thereby ensuring a high signal-to-noise ratio throughout the observation window. Consequently, these differential signals provide a faithful representation of the intrinsic spin evolution as projections of the spin vector $\boldsymbol{S}(t)$ onto the orthogonal axes of the Bloch sphere.

 Figure 2 displays the resulting quantum state tomography. In the circular and diagonal bases (top and middle panels of Fig. 2), pronounced quantum beats are observed. In contrast, $D_{\rm rect}$ remains nearly zero throughout the observation window (bottom of Fig. 2 ). This  vanishingly small amplitude in $D_{\rm rect}$ indicates that the precession axis coincides with the $H/V$ axis. Although minor residual oscillations are present in $D_{\rm rect}$, they can be effectively accounted for by experimental crosstalk from $D_{\rm circ}$ and $D_{\rm diag}$ ($\sim 14\%$ and $\sim 10\%$, respectively). This indicates that the precession axis is well-aligned with the rectilinear basis defined by the excitonic eigenstates $|H\rangle$ and $|V\rangle$, consistent with the AEI-dominated Hamiltonian.

\begin{figure}
\centering
\includegraphics[width=0.9\columnwidth]{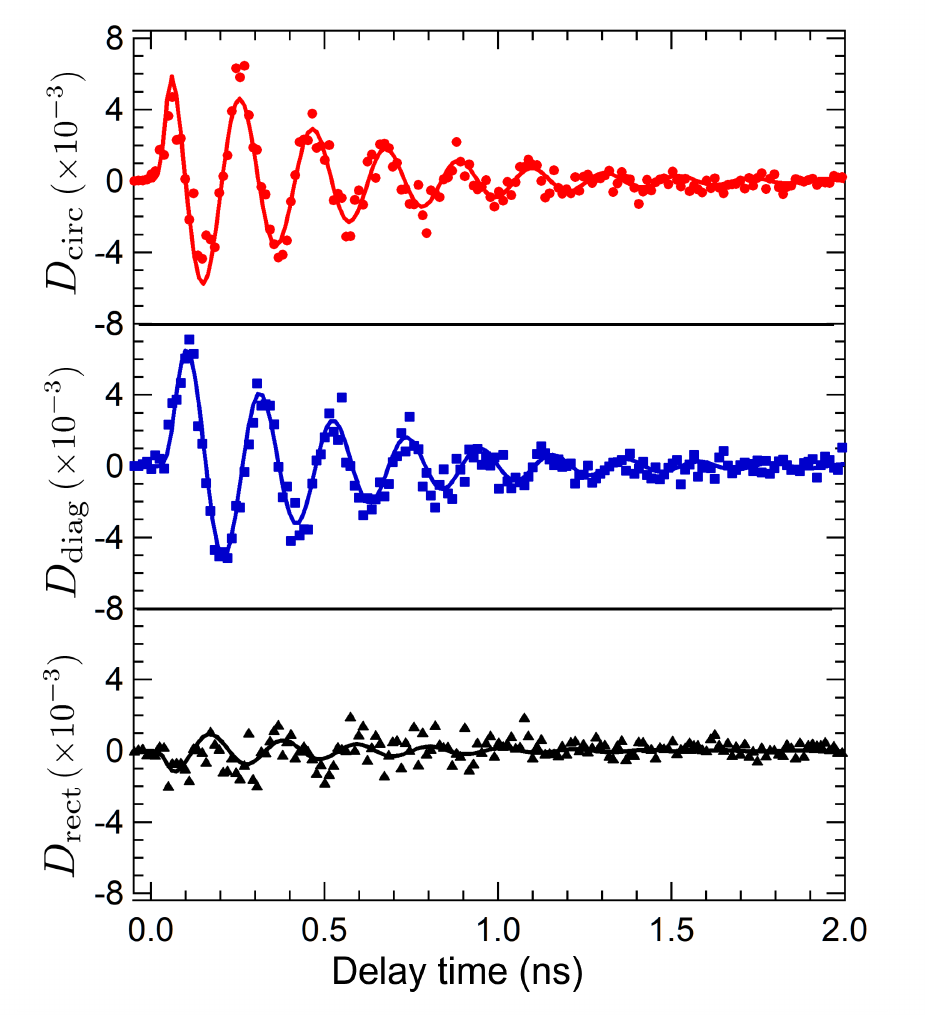}%
\caption{\label{fig2} The differential signals $D_i$ ($i = \text{circ, diag, rect}$), defined as $D_{\rm circ} = \tilde{I}_L - \tilde{I}_R$ (red solid circles), $D_{\rm diag} = \tilde{I}_A - \tilde{I}_D$ (blue solid squares), and $D_{\rm rect} = \tilde{I}_H - \tilde{I}_V$ (black solid triangles), are shown from top to bottom panels. The intensities $I_j$ are normalized by their respective total integrated counts to compensate for relative detection efficiencies. Pronounced quantum beats with a distinct out-of-phase relationship are observed between $D_{\rm circ}$ and $D_{\rm diag}$. In contrast, $D_{\rm rect}$ remains nearly zero throughout the observation window, with its residual oscillation quantitatively accounted for by crosstalk from the other channels. This indicates that the precession axis is locked to the $H/V$ basis,  providing the foundation for the three-dimensional reconstruction of the spin trajectory $\boldsymbol{S}(t)$. Solid curves represent the global fitting results obtained using Eqs. (3)--(4).
}
\end{figure}

To analyze these dynamics quantitatively, we employ a density matrix formalism where the time evolution of the system is governed by the master equation $\frac{d\rho}{dt} = -\frac{i}{\hbar}[\mathcal{H}, \rho] + \mathcal{L}(\rho)$, with the density matrix $\rho$ defined in the circular basis $\{|L\rangle, |R\rangle\}$ including a ground state $|G\rangle$~\cite{Eas13}. The coherent evolution is driven by the Hamiltonian $\mathcal{H} = \frac{1}{2} (\Delta \sigma_x + \delta \sigma_z)$, where $\sigma_x$ and $\sigma_z$ are the Pauli matrices acting on the $\{|L\rangle, |R\rangle\}$ subspace. In this framework, the quantum beat dynamics are governed by the competition between a driving torque $\Delta$ that induce spin precession and a restorative bias $\delta$ along the optical axis. The dissipation is described by the Lindblad term $\mathcal{L}(\rho) = \mathcal{L}_{\tau}(\rho) + \mathcal{L}_{s}(\rho)$. The exciton recombination process to the ground state $|G\rangle$ is given by $\mathcal{L}_{\tau}(\rho) = \sum_{j \in \{L,R\}} (C_{j} \rho C_{j}^\dagger - \frac{1}{2}\{C_{j}^\dagger C_{j}, \rho\})$ with the jump operators $C_{j} = \sqrt{1/\tau}|G\rangle\langle j|$. For the subspace of the excited states $\{|L\rangle, |R\rangle\}$, this term effectively reduces to a uniform decay, $\mathcal{L}_{\tau}(\rho) = -\rho/\tau$, which sets the baseline for the overall signal disappearance.

The second term, $\mathcal{L}_{s}(\rho) = \mathcal{L}_{T_1}(\rho) + \mathcal{L}_{T_2^*}(\rho)$, represents the spin relaxation within the exciton manifold $\{|L\rangle, |R\rangle\}$. Since this process does not involve the ground state $|G\rangle$, the dynamics are described by jump operators acting solely within the excited-state subspace. For the longitudinal spin-flip process ($T_1$), the dissipator is given by the operators $C_{L \to R} = \sqrt{1/T_1}|R\rangle\langle L|$ and $C_{R \to L} = \sqrt{1/T_1}|L\rangle\langle R|$. For the pure dephasing process ($T_2^*$), we employ the operator $C_{ph} = \sqrt{1/(2T_2^*)}(|L\rangle\langle L| - |R\rangle\langle R|)$. These operators lead to the explicit matrix form in the circular basis:$$-\frac{1}{T_1}
\begin{pmatrix}
\rho_{LL}-\rho_{RR} & 0 \\
0 & \rho _{RR}-\rho _{LL}
\end{pmatrix} - \frac{1}{T_2^*}
\begin{pmatrix}
0 & \rho_{LR} \\
\rho_{RL} & 0\end{pmatrix}.$$ Here, $\rho_{ij}$ are the elements of $\rho$ in the circular basis. $T_1$ denotes the longitudinal spin-flip time which equalizes the populations $\rho_{LL}$ and $\rho_{RR}$, while $T_{2}^{*}$ denotes the pure dephasing time which leads to the loss of phase coherence between $\rho_{LR}$ and $\rho_{RL}$ without affecting their populations.

In our system, the aforementioned $\Delta$ and $\delta$ are attributed to the FSS and the Overhauser shift, respectively. It is intuitively clear that the FSS acts as a driving torque for spin precession, as it transforms the basis states $\{|L\rangle, |R\rangle\}$ into their coherent superpositions $\{|H\rangle, |V\rangle\}$. Conversely, the Overhauser shift functions as the longitudinal bias that stabilizes the original basis states $\{|L\rangle, |R\rangle\}$ against this precession.

In self-assembled QDs, the hyperfine interaction often leads to a significant $\delta$, particularly under circular excitation where dynamic nuclear polarization can create a robust internal magnetic field~\cite{Brau05,Kaji08,Urb13,Matsusaki18}. Under a zero external magnetic field, this field typically aligns with the optical axis, causing the spin precession axis to deviate from the equatorial plane of the Bloch sphere. In the present study, however, no discernible signatures?such as a persistent vertical offset in $D_{\rm circ}$?are observed beyond the experimental uncertainty. Moreover, the current signal-to-noise ratio is not sufficient to support a rigorous quantitative assessment of such a field. We therefore adopt the approximation $\delta \approx 0$ to avoid unphysical estimations and prevent the model from becoming over-parameterized. This approach ensures that the extracted values for the FSS ($\Delta$) and effective decay constant remain physically grounded.


Under these conditions, the equations for the circular and diagonal polarization components, $D_{\text{circ}}$ and $D_{\text{diag}}$, are expressed as:
\begin{align}\frac{d}{dt} D_{\text{circ}} &= -\frac{\Delta}{\hbar} D_{\text{diag}} - \gamma_1 D_{\text{circ}}, \\
\frac{d}{dt} D_{\text{diag}} &= \frac{\Delta}{\hbar} D_{\text{circ}} - \gamma_2 D_{\text{diag}},
\end{align}
where $D_{\text{circ}}$ and $D_{\text{diag}}$ correspond to the population difference $\rho_{LL} - \rho_{RR}$ and the imaginary part of the coherence $2\text{Im}[\rho_{LR}]$ that drives the spin precession, respectively. The relaxation rates are given by $\gamma_1 = 1/\tau + 2/T_1$ and $\gamma_2 = 1/\tau+1/T_2^{*}$. These coupled equations lead to a second-order differential equation for $D_{\text{circ}}$:$$\frac{d^2}{dt^2} D_{\text{circ}} + (\gamma_1 + \gamma_2) \frac{d}{dt} D_{\text{circ}} + (\Omega^2 + \gamma_1 \gamma_2) D_{\text{circ}} = 0,$$where $\Omega = \Delta / \hbar$.

In the weak damping limit ($\Omega \gg \gamma_{1,2}$), the solutions for the differential signals are expressed as:
\begin{align}
D_{\text{circ}} &= K_0  \exp(-t/\tau_{\text{eff}}) \cos(\Omega t), \label{eq:D_circ} \\
D_{\text{diag}} &= K_0  \exp(-t/\tau_{\text{eff}}) \sin(\Omega t), \label{eq:D_diag}
\end{align}
where $K_{0}$ is the coherent oscillation amplitude. The effective decay rate is defined as $1/\tau_{\text{eff}} = (\gamma_1 + \gamma_2)/2$. The experimental data are consistently accounted for by a global fit using the model convolved with the IRF. The fitting yields $\Delta = 19.59 \pm 0.08~\mu\text{eV}$, $K_{0}=(7.9\pm0.3) \times 10^{-3}$, and $\tau_{\text{eff}} = 460 \pm 20$~ps, respectively. Notably,  the characteristic $90^\circ$ phase shift was captured without any arbitrary time-zero adjustment, reflecting the near-instantaneous spin formation and the high temporal precision of our measument. From these results, an intrinsic spin relaxation time of $\tau_s \approx 1.15$ ns can be estimated using the relation $1/\tau_{\text{eff}} = 1/\tau + 1/\tau_s$, where $\tau \approx 767$ ps is the exciton lifetime deduced from the TRPL signal as shown in Fig. 1(c). Within our density matrix framework, this relaxation time is expressed as $\tau_s = 2T_1 T_2^* / (T_{1}+2T_2^{*})$. It is important to emohasize that $\tau_s$ significantly exceeds the exciton lifetime, indicating that the spin coherence is remarkably robust and persists throughout the entire emission process. This implies that the exciton spin remains nearly phase-consistent throughout its lifetime, a feature that is further evidenced by the full tomographic reconstruction described below.

By comparing the coherent oscillation amplitude $K_0 = (7.9 \pm 0.3) \times 10^{-3}$ with the previously determined initial population amplitude $f_0 = (28.6 \pm 0.2) \times 10^{-3}$ [Fig. 1(c)], we estimated the initial circular polarization to be $P_0 = K_0 / f_0 \approx 0.28$. This value is physically consistent given that our 2--3 ps excitation pulse provides an energy uncertainty ($\Delta E \approx 150$--$220$ $\mu$eV) significantly exceeding the FSS ($\sim$15--30 $\mu$eV). This ensures the simultaneous and coherent excitation of the $|H\rangle$ and $|V\rangle$ eigenstates, directing the spin initialization toward the $|L\rangle$ state. While $P_0$ appears modest, it predominantly reflects the impact of the extremely fast AEI-induced precession convolved with the finite IRF, leading to a time-averaged initial polarization rather than a lack of deterministic spin injection. While the preceding estimation based on $\tau_{\text{eff}}$ suggests a robust spin lifetime, the intrinsic coherence is more directly and rigorously visualized through three-dimensional tomographic reconstruction, which isolates the spin vector dynamics from the total intensity decay.

\begin{figure}
\centering
\includegraphics[width=0.9\columnwidth]{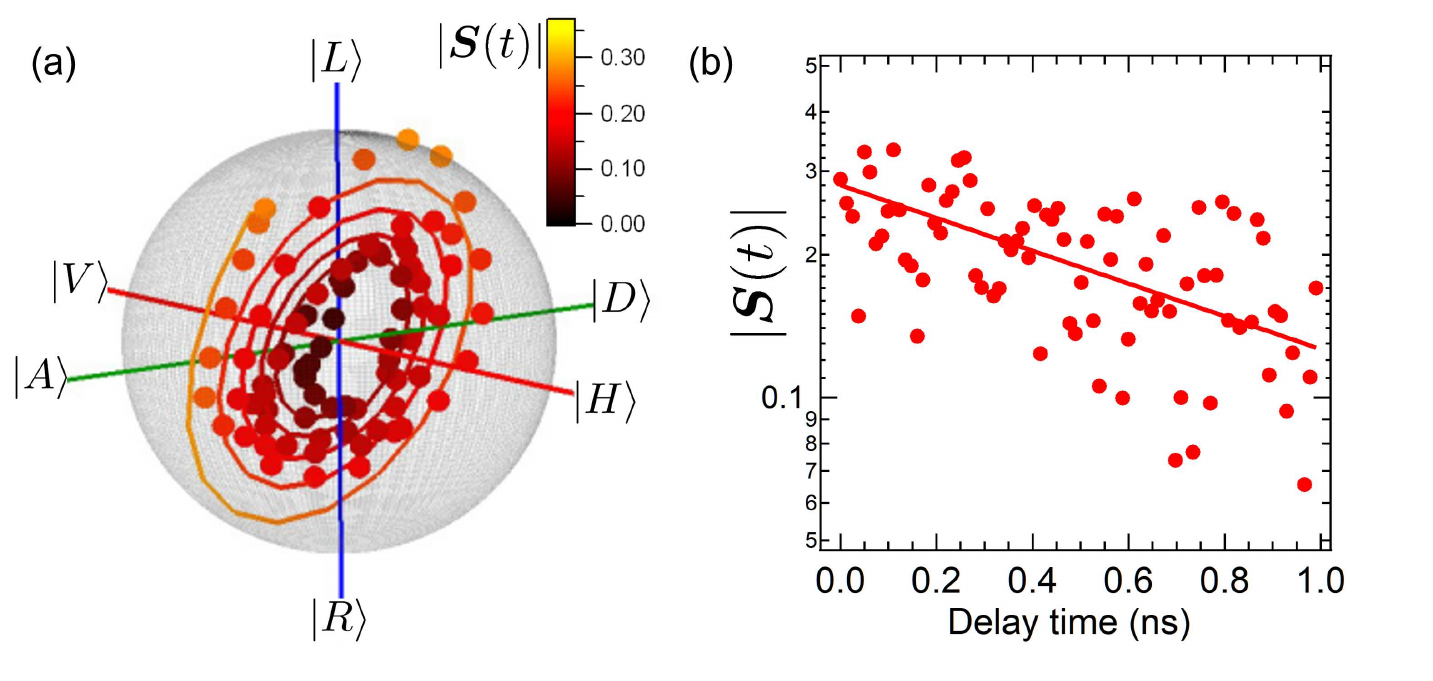}%
\caption{\label{fig3} Three-dimensional reconstruction of the spin polarization vector $\boldsymbol{S}(t)$. (a) Trajectory of $\boldsymbol{S}(t)$ on the Bloch  sphere. The gray sphere shell represents the initial degree of polarization $P_0$, providing a reference for the maximum achievable spin magnitude. The color of data points reflects $|\boldsymbol{S}(t)|$, as indicated by the color scale. The solid curve represents the trajectory calculated from the global fit (including the IRF) as a guide for the eye. To clearly visualize the precession geometry, the plot starts from the first maximum of $D_{\rm circ}$, thereby excluding the initial rise where the vector expands from the origin. (b) Time evolution of the spin magnitude $|\boldsymbol{S}(t)|$. The data are plotted from 0 to 1.0 ns to provide a comprehensive view of the decay, utilizing the normalization method described in the text to avoid noise amplification in the low-count regime. The solid line represents and exponential decay fit, yielding an observed dephasing time of $\tau_{s} = 1.1 \pm 0.2$ ns.
}
\end{figure}

To provide a comprehensive visualization, we reconstructed the three-dimensional spin trajectory $\boldsymbol{S}(t)$ on the Bloch sphere [Fig. 3(a)]. The trajectory exhibits a clear spiral motion starting from $P_0$ and converging toward the center, highlighting the deterministic spin precession. Notably, the observed counter-clockwise rotation ($|L\rangle \to |A\rangle \to |R\rangle \to |D\rangle$) is consistent with the sign of $\Delta(=E_{V}-E_H > 0)$ as defined in Fig. 1(b), where $E_{H/V}$ represents the eigenenergy of $X^{0}$ for $|H/V\rangle$ basis. Furthermore, precession occurs predominantly within the $|L\rangle$--$|R\rangle$ and $|D\rangle$--$|A\rangle$ plane (corresponding to the $D_{\rm circ}$--$D_{\rm diag}$ plane),
 consistent with the observed $90^\circ$ phase shift between $D_{\rm circ}$ and $D_{\rm diag}$ and the negligible oscillation amplitude along the $|H\rangle$--$|V\rangle$ axis ($D_{\rm rect}$). This planar rotation confirms that the spin quantization axis remains well-defined along the optical axis throughout the evolution. 

To quantify the intrinsic dephasing independently of these oscillations, we analyzed the magnitude of the spin polarization vector, 
\begin {align}
|\boldsymbol{S}(t)| &=\frac {\sqrt{D_{\rm rect}^2 + D_{\rm diag}^{2} + D_{\rm circ}^{2}}}{\tilde{I}_{L}+\tilde{I}_{R}}.\end {align}
As shown in Fig. 3(b), $|\boldsymbol{S}(t)|$ exhibits a single exponential decay with a time constant of $\tau_s = 1.1 \pm 0.2$ ns. This value, obtained by fitting beyond the initial rise to avoid IRF-related artifacts, reflects the intrinsic spin coherence of the single exciton. Since $|\boldsymbol{S}(t)|$ is normalized by the total PL intensity, it effectively decouples the spin dephasing from the exciton recombination process, $\mathcal{L}_{\tau}$. Consistently, this direct measurement of $\tau_s$ is in excellent agreement with the value estimated from the individual projections ($\tau_s \approx 1.15$ ns), which was derived from $\tau_{\text{eff}}$ and $\tau$ as discussed earlier. This consistency validates both our tomographic reconstruction and the underlying density matrix model. 

In the limit of long $T_1$, this observed time constant relates to the transverse relaxation time as $\tau_s \approx 2T_2^*$, yielding a pure dephasing time of $T_2^* \approx 550$ ps. The fact that $T_2^*$ is comparable to the exciton lifetime $\tau$ indicates that the exciton spin remains robust against dephasing. The spiral trajectory in Fig. 3(a) visually encapsulates this physics: the radius shrinks at a slower rate than the overall emission decay, while maintaining constant frequency and phase. These results demonstrate the precision of our tomographic reconstruction and the feasibility of stable spin control in these nanostructures.

\section {Summary}
We have performed full quantum state tomography of a single neutral exciton spin in an InAlAs QD under two-LO-phonon quasi-resonant excitation. By analyzing the time-resolved polarization components in three orthogonal bases, we observed pronounced coherent quantum beats that are precisely described by a unified density matrix model based on the anisotropic exchange interaction ($\Delta = 19.59 \pm 0.08 \ \mu\text{eV}$). Our global fitting procedure revealed that the spin-formation process is near-instantaneous (limited by the IRF) and the precession axis is rigidly locked to the $H/V$ basis, indicating a negligible Overhauser shift.

The three-dimensional reconstruction of the spin trajectory $\boldsymbol{S}(t)$ allowed for the direct observation of the intrinsic dephasing process. While the individual polarization projections decay with an effective time constant of $\tau_{\rm eff} \approx 460$ ps, the magnitude of the spin vector $|\boldsymbol{S}(t)|$ reveals a significantly longer observed decay time of $\tau _{s} \approx 1.1$ ns. Significantly, this intrinsic spin relaxation time exceeds the exciton lifetime ($\tau \approx 767$ ps), demonstrating that the spin coherence is preserved throughout the entire exciton recombination process. Even in the limit of long $T_1$, the deduced pure dephasing time $T_2^{*} \approx 550$ ps remains comparable to $\tau$, further highlighting the remarkable robustness of the spin state. 

These results provide a consistent benchmark for the well-controlled initialization and precise tomographic reconstruction of spin qubits in semiconductor nanostructures, underscoring the potential of quasi-resonant excitation for high-fidelity quantum state control for future quantum information applications.


\begin{references}
\bibitem{book1} D. Awschalom, D. Loss, and N. Samarth, \textit{Semiconductor spintronics and quantum computation} (Springer Science \& Business Media, 2002).

\bibitem{shie07} A. J. Shields, Semiconductor quantum light sources, Nat. Photonics \textbf{1} 215 (2007).

\bibitem{Press08}D. Press, T. D. Ladd, B. Zhang, and Y. Yamamoto, Complete quantum control of a single quantum dot spin using ultrafast optical pulses, Nature \textbf{456} 218 (2008).

\bibitem{Ber08}J. Berezovsky, M. H. Mikkelsen, N. G. Stoltz, L. A. Coldren, and D. D. Awschalom, Picosecond coherent optical manipulation of a single electron spin in a quantum dot, Science \textbf{320} 349 (2008).

\bibitem{DeGreve12} K. De Greve, L. Yu, P. L. McMahon, J. S. Pelc, C. Natarajan, N. Y. Kim, E. Abe, S. Maier, C. Schneider, M. Kamp, S. H{\"o}fling, R. H. Hadfield, M. M. Fejer, and Y. Yamamoto, Quantum-dot spin photon entanglement via frequency downconversion to telecom wavelength, Nature \textbf{491}, 421 (2012).

\bibitem{Lodahl15} P. Lodahl, S. Mahmoodian, and S. Stobbe, Interfacing single photons and single quantum dots with photonic nanostructures, Rev. Mod. Phys. \textbf{87}, 347 (2015).

 \bibitem{Bayer02} M. Bayer, G. Ortner, O. Stern, A. Kuther, A. A. Gorbunov, A. Forchel, P. Hawrylak, S. Fafard, K. Hinzer, T. L. Reinecke, S. N. Walck, J. P. Reithmaier, F. Klopf, and F. Sch{\"a}fer, Fine structure of neutral and charged excitons in self-assembled In(Ga)As/(Al)GaAs quantum dots, Phys. Rev. B \textbf{65}, 195315 (2002).

\bibitem{Gammon96} D. Gammon, E. S. Snow, B. V. Shanabrook, D. S. Katzer, and D. Park, Fine structure splitting in the optical spectra of single GaAs quantum dots, Phys. Rev. Lett. \textbf{76}, 3005 (1996).

\bibitem{Tar04} A. I. Tartakovskii, M. N. Makhonin, I. R. Sellers, J. Cahill, A. D. Andreev, D. M. Whittaker, J-P. R. Wells, A. M. Fox, D. J. Mowbray, M. S. Skolnick, K. M. Groom, M. J. Steer, H. Y. Liu, and M. Hopkinson, Effect of thermal annealing and strain engineering on the fine structure of quantum dot excitons, Phys. Rev. B \textbf{70}, 193393 (2004).


\bibitem{Ivchenko97} E. L. Ivchenko, Fine structure of excitonic levels in semiconductor nanostructures, Phys. Status Solidi (a) \textbf{164}, 487 (1997).

\bibitem{Flis01}T. Flissikowski, A. Hundt, M. Lowisch, M. Rebe, and F. Henneberger, Photon beats from a single semiconductor quantum dot, Phys. Rev. Lett. \textbf{86}, 3172 (2001).

\bibitem{Senes05} M. S{\'e}n{\`e}s, B. Urbaszek, X. Marie, T. Amand, J. Tribollet, F. Bernardot, C. Testelin, M. Chamarro, and J.-M. G{\'e}rard, Exciton spin manipulation in InAs/GaAs quantum dots: Exchange interaction and magnetic field effects, Phys. Rev. B \textbf{71}, 115334 (2005).

\bibitem{Carl17} C. Ant{\'on}, P. Hilaire, C. A. Kessler, J. Demory, C. G{\'o}mez, A. Lema{\^i}tre, I. Sagnes, N. D. Lanzillotti-Kimura, O. Krebs, N. Somaschi, P. Senellart, and L. Lanco, Tomography of the optical polarization rotation induced by a single quantum dot in a cavity Optica \textbf{4}, 1326 (2017).

\bibitem {Cog20} D. Cogan, G. Peniakov Z-E. Su, and D. Gershoni, Complete state tomography of a quantum dot spin qubit, Phys. Rev. B \textbf{101}, 035424 (2020).

\bibitem{Brau05} P.-F. Braun, X. Marie, L. Lombez, B. Urbaszek, T. Amand, P. Renucci, V. K. Kalevich, K. V. Kavokin, O. Krebs, P. Voisin, and Y. Masumoto, Direct observation of the electron spin relaxation induced by nuclei in quantum dots, Phys. Rev. Lett. \textbf{94}, 116601 (2005).



\bibitem{Urb13}B. Urbaszek, X. Marie, T. Amand, O. Krebs, P. Voisin, P. Malentinsky, A. H\"{o}gele, and A. Imamoglu, Nuclear spin physics in quantum dots: An optical investigation, Rev. Mod. Phys. \textbf{85}, 79 (2013).

\bibitem {Kik98} J. M. Kikkawa and D. D. Awschalom, Resonant Spin Amplification in $n$-Type GaAs, Phys. Rev. Lett. \textbf{80}, 4313 (1998). 



\bibitem{Kosa08} H. Kosaka, H. Shigyou, Y. Mitsumori, Y. Rikitake, H. Imamura, T. Kutsuwa, K. Arai, and K. Edamatsu, Coherent Transfer of Light Polarization to Electron Spins in a Semiconductor, Phys. Rev. Lett. \textbf{100}, 096602 (2008).

\bibitem{Gio25} G. Peniakov, I. M. Michl, M. Helal, R. joos, M. Jetter, S. L. Portalupl, P. Michler, S. H{\"o}fling, and T. Huber-Loyola, Initialization of neutral and charged exciton spin states in a telecom-emitting quantum dot, Phys. Rev. B \textbf{112}, 085422 (2025).

\bibitem{Sa04} H. Sasakura, S. Adachi, S. Muto, H.-Z. Song, T. Miyazawa, and T. Usuki, Spin depolarization via tunneling effects in asymmetric double quantum dot structure, Jpn. J. Appl. Phys., Part 1 \textbf{43}, 2110 (2004).

\bibitem{Ad07}S. Adachi, N. Yatsu, R. Kaji, S. Muto, and H. Sasakura, Decoherence of exciton complexes in single InAlAs quantum dots measured by Fourier spectroscopy, Appl. Phys. Lett. \textbf{91}, 161910 (2007).


\bibitem{Eas13}P. R. Eastham, A. O. Spracklen, and J. Keeling, Lindblad theory of dynamical decoherence of quantum-dot excitons, Phys. Rev. B \textbf{87}, 195306 (2013).
\bibitem{Kaji08} R. Kaji, S. Adachi, H. Sasakura, and S. Muto, Hysteretic response of the electron-nuclear spin system in single InAlAs quantum dots: Dependences on excitation power and polarization, Phys. Rev. B \textbf{77}, 115345 (2008).

\bibitem{Matsusaki18} R. Matsusaki, R. Kaji, S. Yamamoto, H. Sasakura, and S. Adachi, Quadrupolar effect on nuclear spin depolarization in single self-assembled quantum dots, Appl. Phys. Express \textbf{11}, 085201 (2018).





\end {references}
\end{document}